# Tracking visible pulsed laser annealing of $Hf_{0.5}Zr_{0.5}O_2$ heterostructures with in situ transmission electron microscopy

Aida Amini[1], Shruti Verma[1], Katharina Kohlmann[1], Sebastian Obernberger[1], Jean-Christof Lamanque[1†], Andreas Rüdiger[1], Kenneth R. Beyerlein[1*],

[1]Institut national de la recherche scientifique (INRS), Center of Energy, Materials, and Telecommunications, Varennes, QC, Canada.

* Corresponding author: kenneth.beyerlein@inrs.ca
† Current address: McGill University, Department of Electrical Engineering, Montreal, QC, Canada

**Abstract**:

Laser annealing offers a promising route to investigate and control phase formation in hafnium-zirconium oxide (HZO) thin films under rapid thermal conditions. Due to the wide band gap of this material, previous reports have studied the crystallization of HZO using ultraviolet or infrared light. In contrast, we monitor its crystallization in a $Si_3N_4$/TiN/$Hf_{0.5}Zr_{0.5}O_2$ thin film heterostructure upon irradiation with visible nanosecond laser pulses. This geometry mimics the structure of CMOS devices and harnesses the absorption of TiN in the visible regime to generate the heat necessary for the transformation. Through a series of local in situ measurements using a modified transmission electron microscope, we quantify the relationship between the HZO film thickness, critical laser energy density and the ferroelectric HZO phase fraction, finding a sharp threshold behavior in the laser pulse energy necessary to crystallize HZO. The optimal condition of irradiating an 8-nm HZO film with a single laser pulse with an energy density of 177 mJ/cm² is found to produce 86% of the ferroelectric orthorhombic phase. Heat transfer dynamics within the heterostructure during laser annealing are revealed by finite element simulations, where the partial melting of the silicon nitride substrate is found to play an important role limiting the temperature to 1900 °C. This finding as well as the observed laser pulse energy threshold behavior support a kinetic crystallization pathway involving the tetragonal phase. More generally, these findings provide insight into laser-driven phase engineering in ferroelectric heterostructures and establish a basis for future studies aimed at functional electronic applications.

Following the discovery of ferroelectricity in Si-doped $HfO_2$ films in 2011[1], extensive research has focused on pure and doped $HfO_2$, $ZrO_2$ and their solid solutions, $Hf_xZr_{1-x}O_2$ (HZO)[2-7]. Hafnia-based ferroelectrics have been explored for advanced electronics such as energy conversion and storage, supercapacitors[8,9], non-volatile random-access memories[3,10], ferroelectric field-effect transistors[11,12], ferroelectric tunnel junctions[13-15], and negative capacitance transistors[16]. Compared with perovskite ferroelectrics, fluorite-structured $HfO_2$-based films are compatible with complementary metal-oxide-semiconductor (CMOS) technology, maintain ferroelectricity below 10 nm[17,18], and have lower toxicity. HZO ferroelectrics also integrate well in advanced electronics and memory devices[1,4,5,19-21], and can exhibit robust ferroelectricity in only a few layers[22]. With $Hf_{0.5}Zr_{0.5}O_2$ being the most widely studied composition[23,24], these materials are promising candidates for industrial electronic devices[25].

The stability and formation mechanisms of ferroelectric $HfO_2$-based thin films are still under investigation, and stabilizing the suitable phase for practical use remains an outstanding challenge[4]. This family of materials exhibits several thermodynamically stable polymorphs at different temperature and pressure conditions: monoclinic $P2_1/c$ phase (m-HZO) is stable at room temperature, tetragonal P42/nmc (t-HZO) becomes stable at 1,700 °C; and cubic Fm3m (c-HZO) becomes stable close to 2,500 °C[24]. However, ferroelectricity has been shown to originate from a non-centrosymmetric orthorhombic $Pca2_1$ phase (oIV-HZO)[24,26], which is metastable at room temperature and ambient pressure. Proposed stabilization mechanisms include the surface energy contributions, vacancies, or strain[27]. Understanding these stability factors necessitates a closer examination of the conditions underlying the phase transitions in this polymorphic material.

Various factors such as film thickness, stress, capping layer, and the type of dopants used, influence the ferroelectric oIV-HZO phase formation[28-30]. Specifically, heat treatments with a high heating and cooling rate such as rapid thermal annealing (RTA) has been shown to be critical for oIV-HZO phase crystallization[31-37]. While the heating rate during RTA is regulated by the applied power, the cooling process is limited to ~100 °C/s by radiation and experimental conditions[33]. RTA temperatures between 500 - 1000 °C have been shown to be effective to crystallize hafnia films of depending on the composition and deposition method[38]. Meanwhile, light-based annealing approaches[39-48], have recently demonstrated advantages towards forming oIV-HZO. Among those methods, pulsed laser annealing (PLA) is a promising route for annealing ferroelectric $HfO_2$-based films because it is scalable and enables a low thermal budget industrial manufacturing of electronic devices. It can reach very high surface heating/cooling rates ($10^6$-$10^{10}$ °C/s)[43,47] and high throughput, while control of pulse fluence, spot size, and beam shape enables localized tuning of crystallinity, dopant activation, and surface morphology without affecting adjacent regions. Only a few studies have examined PLA-driven ferroelectric phase formation in hafnia-based films, including the role of HZO composition[47], pulse count[45,47] top electrode effects[44], using KrF (248 nm)[46,47], XeCl (308 nm)[42,45,50], UV (300-400 nm)[44], and Nd:YAG (1064 nm)[27,43].

In this study, we perform *in-situ* PLA on $Si_3N_4$/TiN/$Hf_{0.5}Zr_{0.5}O_2$ thin film heterostructures using a transmission electron microscope (TEM) modified for optical illumination of the sample. This setup allowed us to directly observe and track local changes in the laser-affected region pulse-by-pulse manner. The heterostructure geometry was chosen for a few reasons. TiN is often used as a common conductive diffusion barrier to connect metals and semiconductor layers in microelectronics. Meanwhile, it exhibits a modest absorptivity in the visible regime. We have benefited from this property to use TiN as a heat source in our heterostructure and to explore, from a fundamental perspective, visible laser PLA of a wide-band-gap material such as HZO ($E_g$ = 5.4eV). By separating the light absorber from the active material, photochemical changes induced in the HZO film by the laser light are avoided, which allows for the study of crystallization driven purely by heat.

Laser-intensity titration and single-shot in situ PLA experiments were performed on heterostructures with HZO thicknesses of 7, 8, and 15 nm (see supplementary information for synthesis and experimental details). SAED patterns were acquired from the irradiated region before and after irradiation with each laser pulse to monitor HZO crystallinity. Representative SAED patterns for the 8-nm sample are shown in Fig. 1(a–c), while a full series is presented in Fig. S3. The as-deposited SAED (Fig. 1a) exhibits weak rings on a strong diffuse background dominated by the amorphous $Si_3N_4$ substrate, indicating partial crystallinity of the HZO and TiN. Laser power titration experiments found the sudden appearance of sharp rings corresponding to crystalline HZO upon irradiation with pulse fluence of 177 mJ/cm² (Fig. 1b). Meanwhile, a lower fluence of 153 mJ/cm² was found to crystalize the sample in single-shot measurements. Below this threshold, SAED showed no observable change even after >100 pulses. Comparable SAED patterns were found in both cases (Fig. 1b and 1c), suggesting that the same HZO phase composition results from titration and single-shot PLA experiments. Figure 1(d) shows a TEM image at the boundary of the laser affected region of the 8-nm sample. The laser crystallized region reveals a larger, more granular microstructure than the non-crystalline unaffected area. Single shot and laser titration experiments were also conducted on heterostructures with 7-nm and 15-nm HZO film thickness and the same threshold behavior was found. Unlike the thinner films, the as-deposited 15-nm HZO sample was found to be partially crystalline as well-defined SAED rings belonging to the m-HZO phase are observed in Fig. 1e, indicating that the condition of this thicker sample is closer to bulk-like HZO. Nonetheless, its crystallinity still increased after laser irradiation fluence of 183 mJ/cm² in titration PLA, as indicated by more pronounced rings in Figure 1(f).

Figure 2 plots the crystallization threshold fluence vs HZO thickness for titration and single-shot PLA experiments. It is seen that the threshold slightly increases with thickness; in titration, the 7-nm, 8-nm, and 15-nm films crystallized at 149, 177, and 183 mJ/cm$^2$, and in single-shot, at 122, 153, and 157 mJ/cm$^2$ respectively. Threshold values carry ~2% uncertainty because of laser stability and values found in repeated measurements. This trend likely reflects the slightly larger heated volume, and therefore higher total energy requirement, associated with thicker HZO films in the laser-affected region. A higher threshold fluence observed in the titration experiments than in the single-shot experiments is explained by pre-annealing of TiN by sub-threshold pulses that slightly change its reflectivity and absorption. Since TiN optical properties are known to depend on crystallinity and temperature[51], this pre-annealing can reduce photothermal efficiency, requiring higher fluence to reach the HZO crystallization temperature.

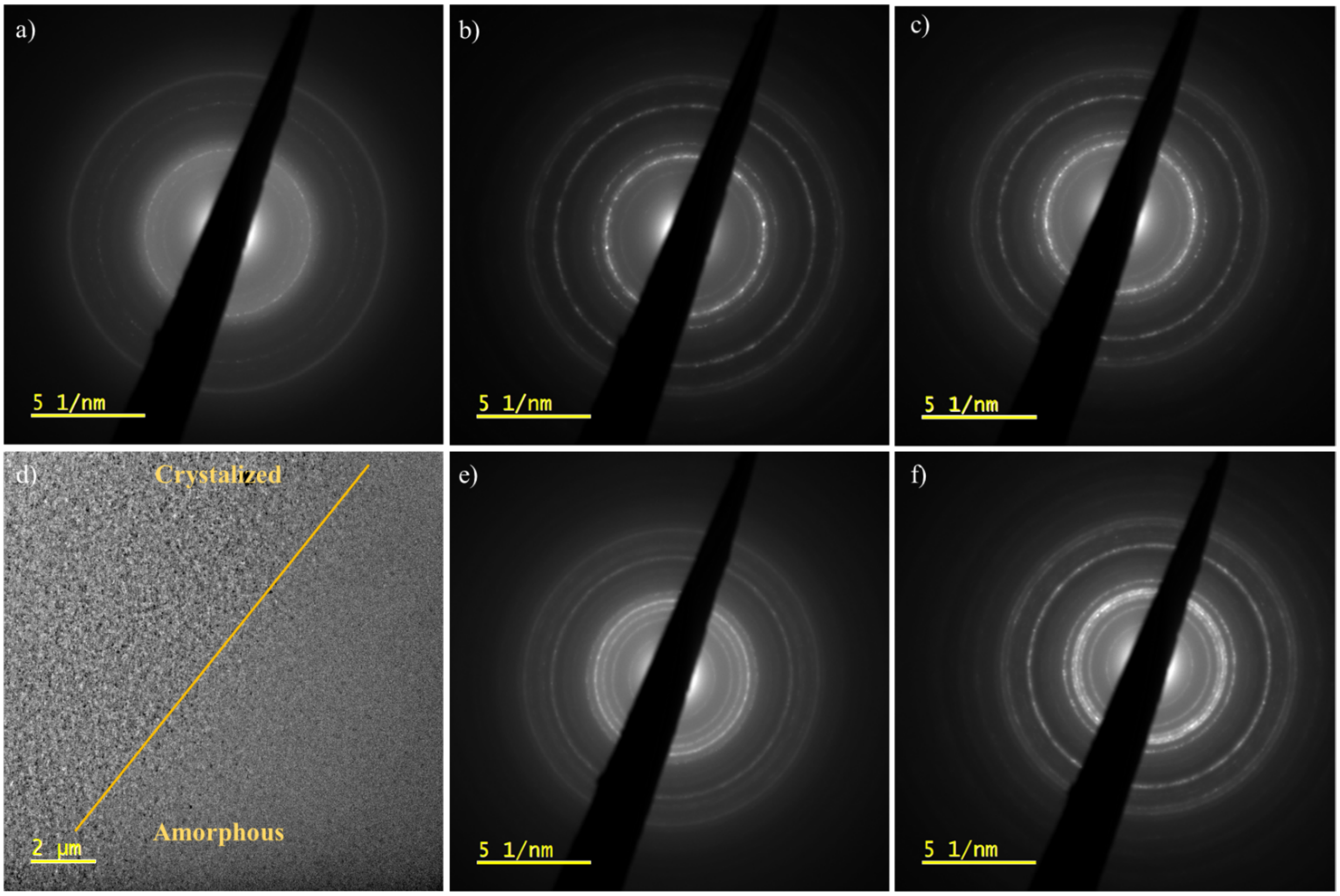

**Figure 1.** SAED images of 8-nm-thick HZO heterostructure (a) before, (b) after laser titration PLA at 177 mJ/cm$^2$, and (c) after single-shot PLA at 153 mJ/cm$^2$. (d) TEM image of the HZO structure showing the boundary of the crystallized region as indicated on the image. SAED pattern of 15-nm-thick HZO heterostructure (e) before and (f) after laser titration PLA at 183 mJ/cm$^2$.

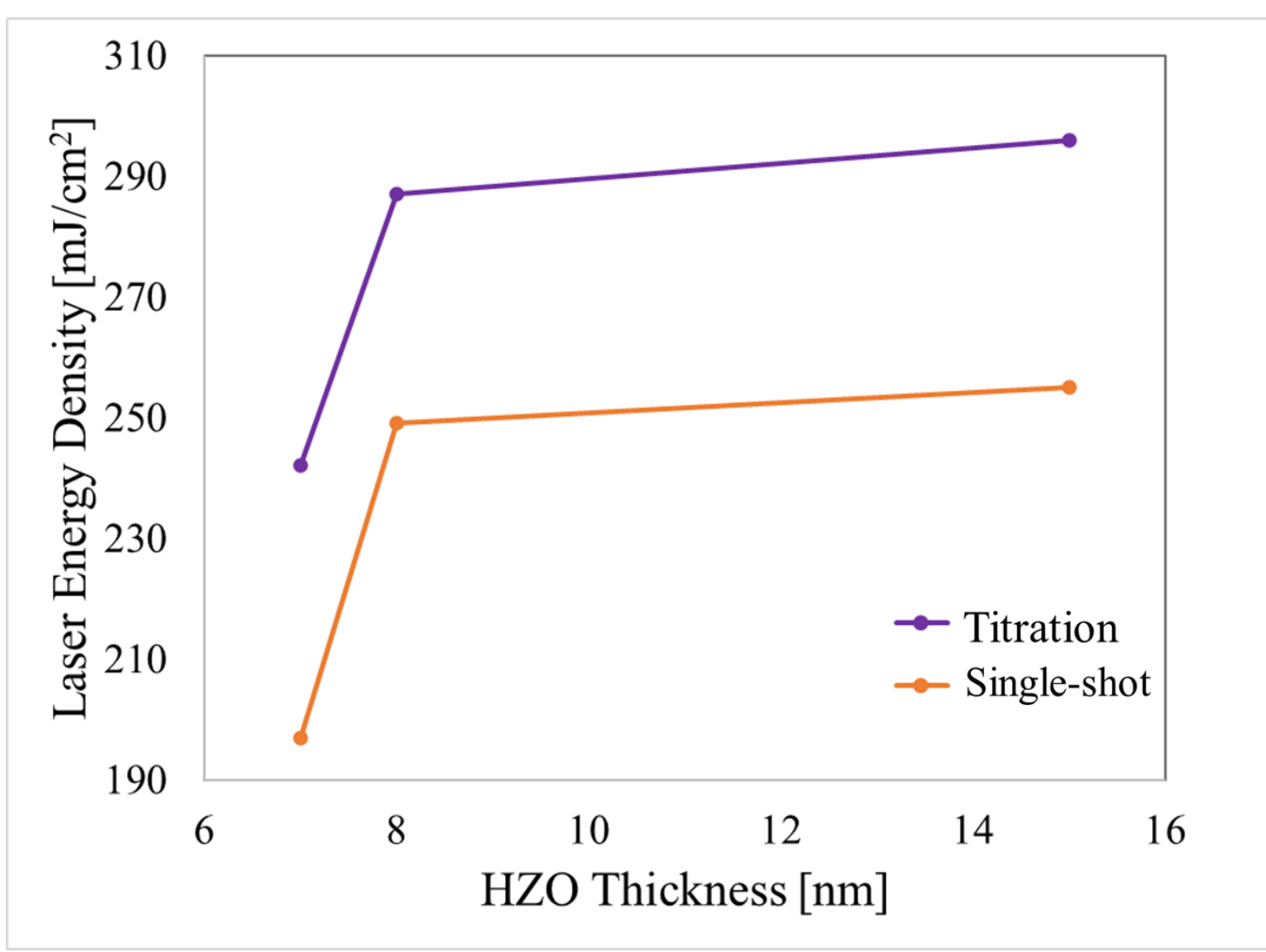

**Figure 2.** Threshold laser fluence for HZO film crystallization as a function of HZO film thickness and PLA experiment type.

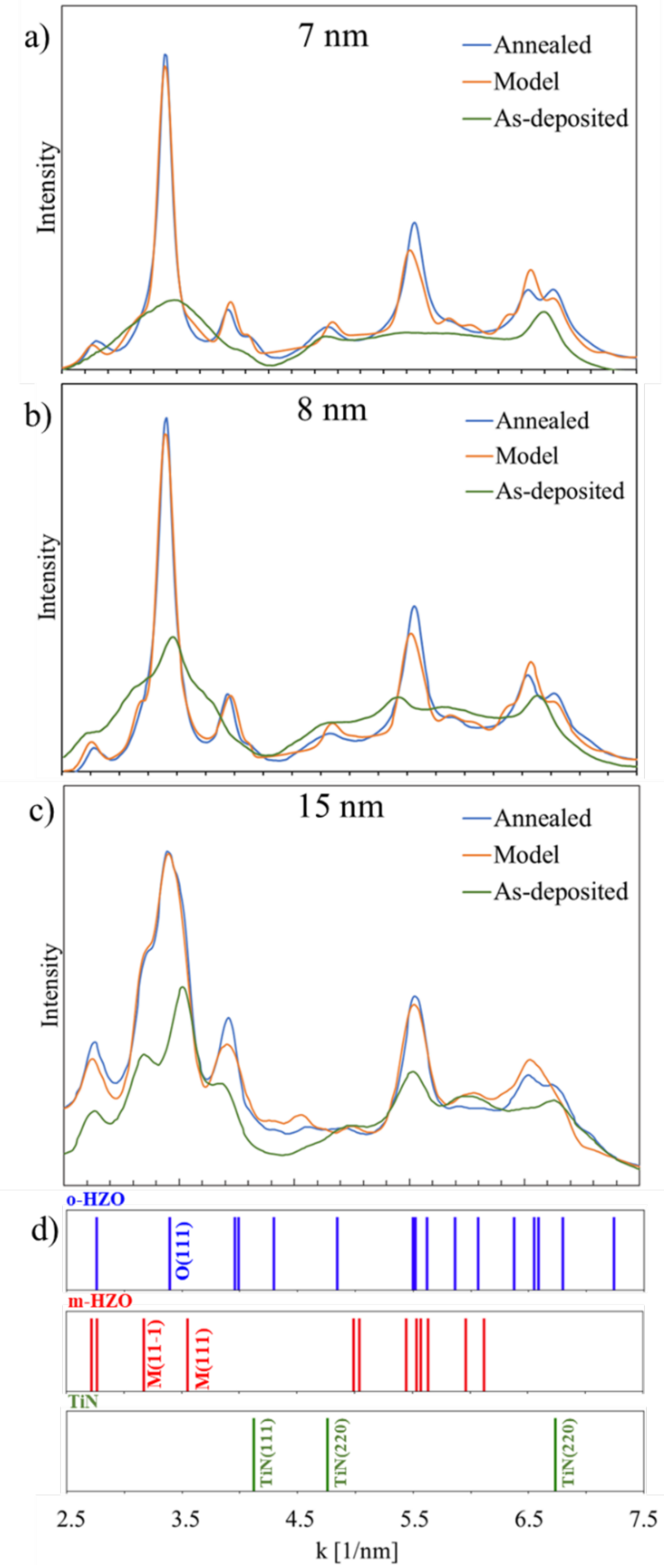

**Figure 3.** Measured SAED radial distribution intensity profiles and best-fit model patterns are shown for the (a) 7-nm ($\chi^2$ = 0.032), (b) 8-nm ($\chi^2$ = 0.021), (c) 15-nm ($\chi^2$ = 0.060) HZO heterostructure samples. As-deposited and annealed profiles correspond to measurements made on the same area of the sample before and after laser intensity titration PLA experiments. Model patterns are associated with the QPA model consisting of oIV-HZO, m-HZO and TiN phases that was found to best fit annealed patterns. (d) The positions of dominant peaks from each phase used in the model are indicated by vertical lines.

Intensity profiles versus scattering vector magnitude ($k$) were then calculated by azimuthal averaging of the measured SAED images. Figures 3(a)-(c) compares the 7, 8, and 15-nm-thick HZO samples as-deposited and after laser-titration crystallization, with predominant positions of oIV-HZO, m-HZO and TiN peaks positions marked in Figure 3(d). Notably, broad peaks at $k$ = 3.18 and 3.55 $nm^{-1}$ are evident in all as-deposited traces (green lines), which agree best with m-HZO $(11\text{-}1)_m$ and $(111)_m$ reflections respectively. This suggests that the m-HZO phase partially crystallizes during the HZO film deposition process. These two peaks are more clearly resolved in in 15-nm-thick sample SAED. After PLA, multiple peaks consistent with oIV-HZO appear, including a major peak at $k$ = 3.41 $nm^{-1}$ corresponding to $(111)_o$. However, this reflection can overlap with the tetragonal phase.[47]

To overcome this ambiguity, we then performed quantitative phase analysis on the annealed sample intensity profiles using the algorithm described in the supplementary information. Models assuming different polymorphic phase compositions of the HZO film combined with the peaks corresponding to TiN and diffuse background were attempted. Only laser titration measurements were analyzed as similar diffraction patterns were measured after single-shot PLA experiments. In each case, only the relative scale factors, peak widths and background functions were refined during the fitting procedure, meaning that neither the relative intensities nor the positions of the diffraction peaks for a given phase were allowed to change. Figure S4 shows a comparison of the measured pattern with best-fit models composed of TiN and either oIV-HZO, m-HZO, or t-HZO phases. The TiN+oIV-HZO model provided the best fit to the experimental data for the 7-nm and 8-nm samples, while the monoclinic and tetragonal models failed to fit the data accurately, with a clear mismatch. Then, the patterns were modelled assuming a mixture of TiN and m-HZO with either the oIV-HZO or the t-HZO phases. The TiN+m-HZO+oIV-HZO model was found to result in the best fit for all HZO thicknesses. Notably, the TiN+m-HZO+t-HZO model could not account for the peak appearing at $k$ = 2.8 $nm^{-1}$, as shown in Figure S4(d). Admittedly, the peak at 5.6 $nm^{-1}$ is found to be in slightly better agreement in this figure as compared to Figure 3(b). However, disagreement in this region of the pattern can also be explained due to background contributions from changes in the $Si_3N_4$ and partially crystalline TiN layers. Therefore, we conclude the ability of the TiN+m-HZO+oIV-HZO model to fit all peaks in the $k$-range of 2.5 to 4.5 $nm^{-1}$ justifies its validity.

The relative scale factors corresponding to the best fit TiN+m-HZO+oIV-HZO model was used to calculate the fraction of each phase present in the laser annealed samples, as described in the supplementary information. In each case, the most abundant phase was found to be the oIV-HZO phase. The oIV-HZO phase fractions for the 7-nm, 8-nm and 15-nm thick HZO samples were found to be 76 ± 2.7 %, 86 ± 1.9 % and 63 ± 1.8 % respectively. The uncertainty in this value was found from propagating the uncertainties in the scale parameters determined from the covariance matrix of the least-square fitting routine. This result suggests that the optimal HZO film thickness to maximize the fraction of ferroelectric oIV-HZO formed by PLA is around 8 nm, which is consistent with other pulsed laser annealing experiments[27]. It is also worth noting that our results show that PLA was able to significantly increase the orthorhombic phase fraction in the 15-nm-thick HZO sample, which showed a predominant m-HZO phase character in its as-deposited state.

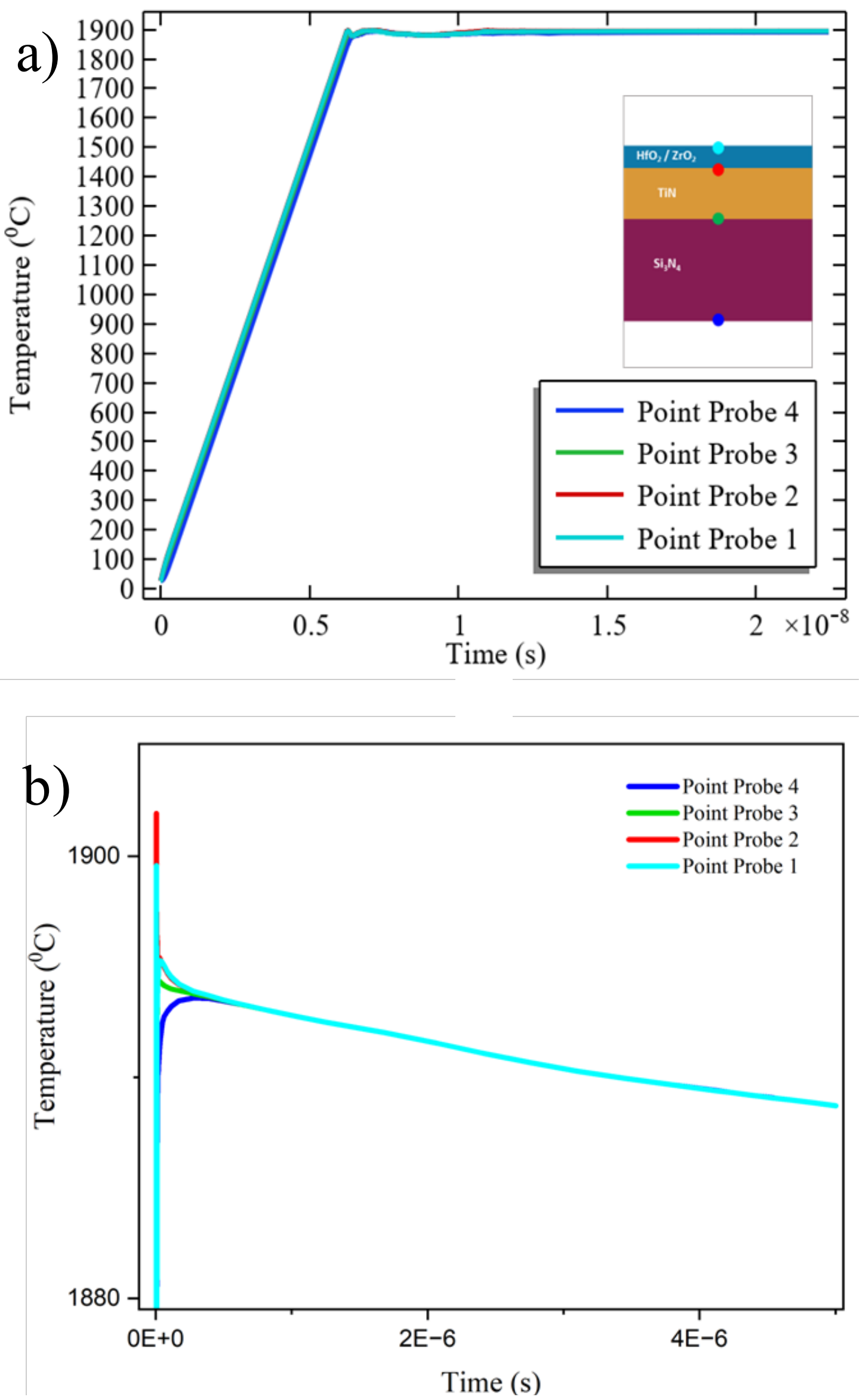

**Figure 4.** (a) Temperature evolution of laser-induced heating in the 8-nm HZO heterostructure probed at different depths. Inset shows the location of the probes. (b) Long-time evolution of temperature showing beginning of cooling.

A finite-element simulation using COMSOL Multiphysics version 6.2 was carried out to gain insight into the temperature evolution in the $Si_3N_4$/TiN/HZO heterostructure upon laser irradiation. The three-dimensional model used the measured stack thicknesses and represented the 11-ns laser pulse as Gaussian spatial, flat-top temporal heat sources in the TiN and HZO layers. Heat was primarily generated in the TiN layer as it has an absorption coefficient of $4.6 \times 10^5$ $cm^{-1}$ at 532 nm[52], while that of hafnia and $Si_3N_4$ are respectively 0.64 $cm^{-1}$ [53] and zero. The lateral edges were fixed at 27 °C and the top and bottom surfaces were thermally insulated to mimic the vacuum environment. A full explanation of the 3D simulation methodology, including heat source expressions, mesh dimensions, vacuum boundary conditions, and the material constants utilized for the HZO layer, is provided in the supplementary information (Figure S5). To accurately bound the temperature evolution, the model incorporates the partial melting of the $Si_3N_4$ substrate at 1900 °C using the apparent heat capacity method.

Upon laser irradiation, the 8-nm HZO heterostructure was found to heat uniformly through its thickness as shown in Figure 4. The rising slope of the temperature increase corresponds a heating rate of $3 \times 10^{11}$ °C/s and a peak temperature near 1900 °C was obtained within 6 ns. These values are is consistent with those reported in other PLA experiments[42,43]. The temperature of the hafnia was found to follow closely that of the TiN, as a temperature difference less than 10 °C through the sample thickness is evident from the slight offset of traces in Fig. 4(a) and different early time dynamics in Fig. 4(b). At 6 ns, as the temperature at the TiN/$Si_3N_4$ interface approaches 1900 °C, heat generated in the TiN layer begins to be consumed by the heat of formation of $Si_3N_4$ melting, causing the temperature of the system to stabilize. Between 6 ns and 11 ns, the TiN continues to absorb light and generate heat, but this energy is absorbed by the $Si_3N_4$ to continue the melt process. At the end of the laser pulse (11 ns), the temperature of the bottom layer is found to remain slightly below its melting point, indicating that not enough heat is generated within by the pulse to completely melt the $Si_3N_4$ layer. This suggests that the silicon nitride layer is partially melted at these incident laser energies, and the endothermic nature of this process serves to limit the peak temperature of the sample close to 1900 °C. Control simulations without this phase change predicted unrealistic temperatures (>4000 °C), and simulations varying the HZO thickness and properties confirmed that HZO thickness variations had negligible impact on this maximum temperature (Figure S6). After the laser pulse, the sample was found to cool at a rate of $-10^6$ °C/s, as found from measuring the slope of the curve in Figure 4(b). This cooling rate is primarily limited by sample thickness and thermal conductivity of the heterostructure materials. At this rate, the sample is expected to reach room temperature within 1.9 ms. This is much longer than the incident laser pulse but is still much shorter than the 30-60 second time scales of RTA heat treatments which are commonly used to crystallize HZO. Simulations of the full cooling process were not performed as they would require an unfeasible computation time.

A kinetic barrier model has been proposed for the oIV-HZO crystallization process involving the formation of a t-HZO precursor at 600-750 °C during RTA[54]. In this model, the t-HZO phase is formed at high temperature and transforms into the metastable oIV-HZO phase after a rapid temperature quench, as the formation of m-HZO is prevented by a high activation barrier. We believe this model also explains our observations of oIV-HZO formation by PLA, even though our simulations show that the sample should reach a higher peak temperature around 1900 °C. At this temperature, t-HZO is still expected to be the most stable phase[24] and the rapid heating will lead to superheating that will in turn result in rapid crystallization of this phase through thickness of the film. As the sample cools below the 1700 °C, which is the lower limit for bulk t-HZO stability, the transformation to the m-HZO phase is again slowed by the large activation energy barrier between the t-HZO and m-HZO phases. Instead, the metastable oIV-HZO phase will form as the sample continues to cool near room temperature. The lack of significant m-HZO phase in our measurements suggests that the quench time of 1.9 ms is sufficiently fast to avoid this transformation and form the oIV-HZO phase. It is also worth noting that our experiments were performed in a high-vacuum environment, which can further enhance the formation of the ferroelectric oIV-HZO phase. Studies have shown that vacuum-annealed HZO can have enhanced ferroelectric properties due to higher concentration of oxygen vacancies, appropriate grain size, and potentially low C–O bonds[55].

In summary, in situ PLA experiments were conducted in a modified TEM on a series of $Si_3N_4$/TiN/HZO thin film heterostructures by irradiation with a nanosecond pulsed visible laser light. Through SAED measurements, we determined that the dominant HZO phase formed in the laser-affected volume is the ferroelectric orthorhombic phase, oIV-HZO. The laser fluence threshold for crystallization was found to slightly increase with the HZO film thickness. Furthermore, using a developed quantitative phase analysis of the SAED profiles, we found the 8-nm-thick HZO film crystallized with 86 ± 1.9 % of the oIV-HZO phase which was the highest of the three thicknesses studied. The quantitative phase analysis ruled out the presence of the t-HZO phase in the sample, which is further support for ferroelectricity in this material originating from the oIV-HZO phase. Finite-element simulations of the peak temperature achieved in the heterostructure are just above the t-HZO phase transition temperature. This supports a kinetic barrier model for oIV-HZO phase crystallization. This work demonstrates the use of visible light absorbing layers as an effective platform for studying rapid annealing of thin films of insulating materials. Decoupling the heat source from the active media allows for more tunability of the laser parameters used in PLA processes. Ongoing research is focused on using this platform to study open questions about the time scales involved in this phase transformation and quench rate dependence on the oIV-HZO phase fraction. This is intended to provide a foundation for future studies of PLA in HZO systems, including eventual assessment of low-thermal-budget processing routes for ferroelectric devices.[55]

## Supplementary Material

Supplementary figures and tables of the experimental configuration, supplementary measurements and simulation details can be accessed using the following link.

## Acknowledgements

We would like to acknowledge the assistance of Patrick Soucy for sample lamella preparation by FIB-SEM and technical support. This work was performed using the Infrastructure for Advanced Imaging instruments, an INRS-EMT facility supported by the Canada Foundation for Innovation (Project 31018) and Ministère de l'Économie et de l'Innovation du Québec. This work was also supported in part by Natural Sciences and Engineering Research Council of Canada (NSERC, RGPIN-2021-03797) and Fonds de Recherche du Québec–Nature et Technologies (ERP-2024-NC-329200). A.R. gratefully acknowledges financial support from NSERC (RGPIN-2024-06730).

## References

1. T. Böscke, J. Müller, D. Bräuhaus, U. Schröder and U. Böttger, *Applied Physics Letters*, **99**, 102903 (2011).
2. J. Müller, P. Polakowski, S. Mueller and T. Mikolajick, *ECS Journal of Solid State Science and Technology*, **4**, N30 (2015).
3. J. Muller, T. S. Boscke, U. Schroder, S. Mueller, D. Brauhaus, U. Bottger, L. Frey and T. Mikolajick, *Nano letters*, **12**, 4318-4323 (2012).

4. M. H. Park, Y. H. Lee, H. J. Kim, Y. J. Kim, T. Moon, K. D. Kim, J. Mueller, A. Kersch, U. Schroeder and T. Mikolajick, *Advanced Materials*, **27**, 1811-1831 (2015).
5. T. Francois, L. Grenouillet, J. Coignus, P. Blaise, C. Carabasse, N. Vaxelaire, T. Magis, F. Aussenac, V. Loup and C. Pellissier, *2019 IEEE International Electron Devices Meeting (IEDM)*, San Francisco, CA, USA, 15.7.1 (2019).
6. M. H. Park, H. J. Kim, Y. J. Kim, W. Lee, T. Moon, K. D. Kim and C. S. Hwang, *Applied Physics Letters*, **105**, 072902 (2014).
7. S. Mueller, J. Mueller, A. Singh, S. Riedel, J. Sundqvist, U. Schroeder and T. Mikolajick, *Advanced Functional Materials*, **22**, 2412-2417 (2012).
8. M. Hoffmann, U. Schroeder, C. Künneth, A. Kersch, S. Starschich, U. Böttger and T. Mikolajick, *Nano Energy*, **18**, 154-164 (2015).
9. T. Mittmann, F. P. Fengler, C. Richter, M. H. Park, T. Mikolajick and U. Schroeder, *Microelectronic Engineering*, **178**, 48-51 (2017).
10. M. Pešić, M. Hoffmann, C. Richter, T. Mikolajick and U. Schroeder, *Advanced Functional Materials*, **26**, 7486-7494 (2016).
11. T. Mikolajick, S. Slesazeck, M. H. Park and U. Schroeder, *Mrs Bulletin*, **43**, 340-346 (2018).
12. E. T. Breyer, H. Mulaosmanovic, T. Mikolajick and S. Slesazeck, *Applied Physics Letters*, **118**, 050501-050501-050501-050507 (2021).
13. L. Chen, T.-Y. Wang, Y.-W. Dai, M.-Y. Cha, H. Zhu, Q.-Q. Sun, S.-J. Ding, P. Zhou, L. Chua and D. W. Zhang, *Nanoscale*, **10**, 15826-15833 (2018).
14. F. Ambriz-Vargas, G. Kolhatkar, M. Broyer, A. Hadj-Youssef, R. Nouar, A. Sarkissian, R. Thomas, C. Gomez-Yáñez, M. A. Gauthier and A. Ruediger, *ACS applied materials & interfaces*, **9**, 13262-13268 (2017).
15. F. Ambriz-Vargas, G. Kolhatkar, R. Thomas, R. Nouar, A. Sarkissian, C. Gomez-Yáñez, M. Gauthier and A. Ruediger, *Applied Physics Letters*, **110**, 093106-093101-093106-093105 (2017).
16. M. Hoffmann and S. Salahuddin, *MRS Bulletin*, **46**, 930-937 (2021).
17. M. H. Park, Y. H. Lee, T. Mikolajick, U. Schroeder and C. S. Hwang, *Mrs Communications*, **8**, 795-808 (2018).
18. Y. Jang, Y. Jeong, D. P. Pham and J. Yi, *Transactions on Electrical and Electronic Materials*, 1-7 (2024).
19. N. Nuraje and K. Su, *Nanoscale*, **5**, 8752-8780 (2013).
20. S. Dünkel, M. Trentzsch, R. Richter, P. Moll, C. Fuchs, O. Gehring, M. Majer, S. Wittek, B. Müller and T. Melde, *IEEE International Electron Devices Meeting (IEDM),* 2017.
21. L. Grenouillet, T. Francois, J. Coignus, S. Kerdiles, N. Vaxelaire, C. Carabasse, F. Mehmood, S. Chevalliez, C. Pellissier and F. Triozon, *IEEE Symposium on VLSI Technology,* 2020.
22. J. Muller, T. S. Boscke, U. Schroder, R. Hoffmann, T. Mikolajick and L. Frey, *IEEE Electron Device Letters*, **33**, 185-187 (2012).
23. S. J. Kim, J. Mohan, S. R. Summerfelt and J. Kim, *Jom*, **71**, 246-255 (2019).
24. U. Schroeder, M. H. Park, T. Mikolajick and C. S. Hwang, *Nature Reviews Materials*, **7**, 653-669 (2022).
25. Z. Zhou, J. Zhou, X. Wang, H. Wang, C. Sun, K. Han, Y. Kang, Z. Zheng, H. Ni and X. Gong, *IEEE Electron Device Letters*, **41**, 1837-1840 (2020).

26. X. Sang, E. D. Grimley, T. Schenk, U. Schroeder and J. M. LeBeau, *Applied Physics Letters*, **106**, 162905-162901-162905-162904 (2015).
27. A. Frechilla, M. Napari, N. Strkalj, E. Barriuso, K. Niang, M. Hellenbrand, P. Strichovanec, F. M. Simanjuntak, G. Antorrena and A. Flewitt, *Applied Materials Today*, **36**, 102033, 102031-102010 (2024).
28. J. Müller, U. Schröder, T. Böscke, I. Müller, U. Böttger, L. Wilde, J. Sundqvist, M. Lemberger, P. Kücher and T. Mikolajick, *Journal of Applied Physics*, **110**, 1141131-1141135 (2011).
29. T. Böscke, S. Teichert, D. Bräuhaus, J. Müller, U. Schröder, U. Böttger and T. Mikolajick, *Applied Physics Letters*, **99**, 1129041-1129043 (2011).
30. E. Yurchuk, J. Müller, S. Knebel, J. Sundqvist, A. P. Graham, T. Melde, U. Schröder and T. Mikolajick, *Thin Solid Films*, **533**, 88-92 (2013).
31. L. Xu, T. Nishimura, S. Shibayama, T. Yajima, S. Migita and A. Toriumi, *Journal of Applied Physics*, **122**, 1241041-1241047 (2017).
32. S. Shibayama, T. Nishimura, S. Migita and A. Toriumi, *Journal of Applied Physics*, **124**, 1841011-1841017 (2018).
33. D. Bouhafs, N. Khelifati, Y. Kouhlane and R. S. Kaddour, *Materials Research Express*, **6**, 055907 055901-055910 (2019).
34. Z. Quan, M. Wang, X. Zhang, H. Liu, W. Zhang and X. Xu, *AIP Advances*, **10**, 085024-085021-085024-085026 (2020).
35. T.-H. Ryu, S.-J. Yoon, S.-Y. Na and S.-M. Yoon, *Current Applied Physics*, **19**, 1383-1390 (2019).
36. J. Wang, D. Wang, Q. Li, A. Zhang, D. Gao, M. Guo, J. Feng, Z. Fan, D. Chen and M. Qin, *IEEE Electron Device Letters*, **40**, 1937-1940 (2019).
37. J. Wu, F. Mo, T. Saraya, T. Hiramoto and M. Kobayashi, *Applied Physics Letters*, **117**, 2529041- 2529045 (2020).
38. H. A. Hsain, Y. Lee, M. Materano, T. Mittmann, A. Payne, T. Mikolajick, U. Schroeder, G. N. Parsons and J. L. Jones, *Journal of Vacuum Science & Technology A*, **40**, 010803 (2022).
39. S. Migita, H. Ota, K. Shibuya, H. Yamada, A. Sawa, T. Matsukawa and A. Toriumi, *Japanese Journal of Applied Physics*, **58**, SBBA07-01-SBBA07-06 (2019).
40. É. O'Connor, M. Halter, F. Eltes, M. Sousa, A. Kellock, S. Abel and J. Fompeyrine, *Apl Materials*, **6**, 1211031-1211037 (2018).
41. H. Tanimura, Y. Ota, H. Kawarazaki, S. Kato and Y. Nara, *Japanese Journal of Applied Physics*, **62**, SC1044 (2023).
42. T. Tabata, *Applied Physics Express*, **13**, 0155091-0155094 (2019).
43. N. Volodina, A. Dmitriyeva, A. Chouprik, E. Gatskevich and A. Zenkevich, *physica status solidi (RRL)–Rapid Research Letters*, **15**, 21000821-21000827 (2021).
44. T. Tabata, S. Halty, F. Rozé, K. Huet and F. Mazzamuto, *Applied Physics Express*, **14**, 115503-115510 (2021).
45. T. Ali, R. Olivo, S. Kerdiles, D. Lehninger, M. Lederer, D. Sourav, A. Royet, A. Sünbül, A. Prabhu and K. Kühnel, *IEEE International Memory Workshop (IMW)*, 2022.
46. A. P. Crema, M. C. Istrate, A. Silva, V. Lenzi, L. Domingues, M. O. Hill, V. S. Teodorescu, C. Ghica, M. J. Gomes and M. Pereira, *Advanced Science*, **10**, 22073901-22073910 (2023).
47. M. S. Song, K. Park, K. Lee, J. W. Cho, T. Y. Lee, J. Park and S. C. Chae, *ACS Applied Electronic Materials*, **5**, 117-122 (2023).

48. S. Kang, W.-S. Jang, A. N. Morozovska, O. Kwon, Y. Jin, Y.-H. Kim, H. Bae, C. Wang, S.-H. Yang and A. Belianinov, *Science*, **376**, 731-738 (2022).
49. T. Tsuchiya, A. Watanabe, Y. Imai, H. Niino, I. Yamaguchi, T. Manabe, T. Kumagai and S. Mizuta, *Japanese journal of applied physics*, **38**, L823 (1999).
51. S. Asgary, Z. Ebrahiminejad and A. H. Ramezani, *Journal of Interfaces, Thin Films, and Low dimensional systems*, **6**, 591-602 (2022).
52. J. Pflüger, J. Fink, W. Weber, K. Bohnen and G. Crecelius, *Physical Review B*, **30**, 1155-1163 (1984).
53. T. J. Bright, J. I. Watjen, Z. Zhang, C. Muratore and A. A. Voevodin, *Thin Solid Films*, **520**, 6793-6802 (2012).
54. M. H. Park, Y. H. Lee, T. Mikolajick, U. Schroeder and C. S. Hwang, *Advanced Electronic Materials*, **5**, 18005221-180052211 (2019).
55. T. Murakami, K.-i. Haga and E. Tokumitsu, *Japanese Journal of Applied Physics*, **59**, SPPB03-01-SPPB03-09 (2020).

# Supplementary Information

## Tracking visible pulsed laser annealing of $Hf_{0.5}Zr_{0.5}O_2$ heterostructures with in situ transmission electron microscopy

Aida Amini[1], Shruti Verma[1], Katharina Kohlmann[1], Sebastian Obernberger[1], Jean-Christof Lamanque[1†], Andreas Rüdiger[1], Kenneth R. Beyerlein[1*],

[1]Institut national de la recherche scientifique (INRS), Center of Energy, Materials, and Telecommunications, Varennes, QC, Canada.

### 1. Experimental Methods

#### 1.1. Sample Preparation

The samples used in this study consisted of heterostructures of $Si_3N_4$/TiN/HZO thin films. These layers were deposited using by magnetron sputtering on a TEM grid with silicon nitride windows. Specifically, a Ted Pella silicon nitride TEM window array (membrane dimensions: 0.10 mm × 0.10 mm, 50 nm thick; frame dimensions: 3 mm diameter, 200 µm thick silicon) was used as the substrate for this work (Figure S1). First, a 23 nm-thick film of TiN was deposited on the grid at 500°C using a Ti target and nitrogen plasma (7 sccm, 40 W). The HZO layer was then deposited on top of the TiN layer at room temperature using an HZO ($Hf_{0.5}Zr_{0.5}$) target and Oxygen+Argon plasma (Ar/$O_2$: 1/2 sccm, 20 W). We obtained HZO film thicknesses of 7, 8, and 15 nm for deposition times of 60, 90, and 180 minutes respectively. These thicknesses were measured from TEM images of cross-section lamellae prepared using a Tescan Lyra 3 focused ion beam – scanning electron microscope (FIB-SEM) system, such as that shown in Figure 1. The platinum layer shown in the figure was deposited for protection during lamella preparation by the FIB and was not present in the PLA experiment samples.

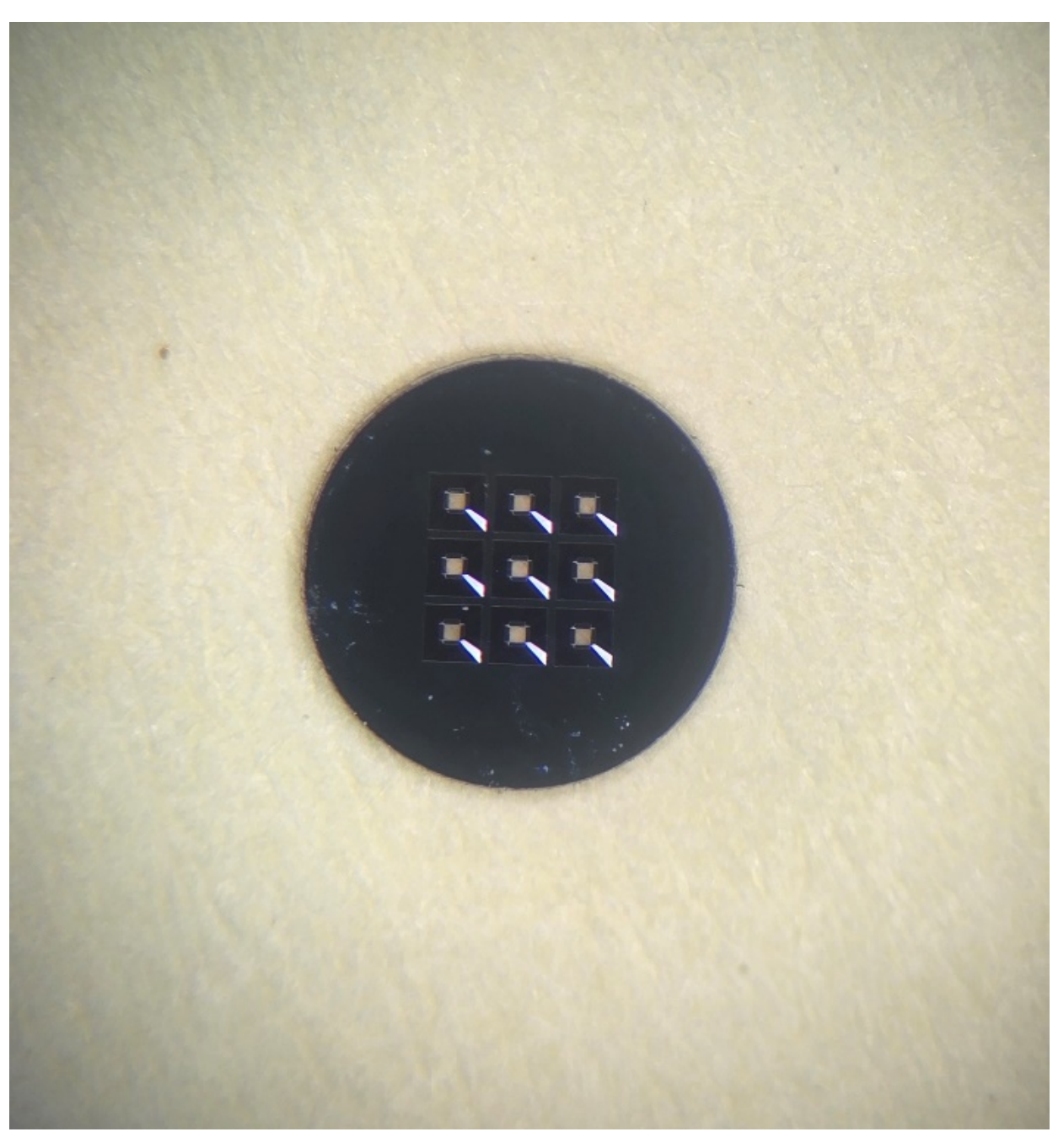

**Figure S1**. Light microscope image of the Ted Pella silicon nitride TEM window array. Membrane dimensions: 0.10 mm × 0.10 mm, 50 nm thick; Frame dimensions: 3 mm diameter, 200 µm thick silicon.

Alt text: Photograph of circular TEM grid that has a 3 by 3 array of 9 square windows.

### 1.2. In situ laser processing

Figure S2 also shows a schematic of the modified TEM at INRS, which was used to perform all *in-situ* PLA experiments. It is based on a JEM-2100 PLUS microscope (JEOL) and equipped with a post-column Gatan Quantum GIF. The column has been modified to allow for a laser beam to be focused on the sample inside the microscope. Images and diffraction patterns were captured in continuous thermionic emission mode with an accelerating voltage of 200 kV. The pump laser system is based on a nanosecond pulsed neodymium-doped yttrium−aluminum−garnet (Nd:YAG) laser (Northrup Grumman). The fundamental wavelength (1064 nm) was frequency doubled to 532 nm using a Type-I barium borate (BBO) crystal. The laser was operated at a repetition rate of 10 Hz and produced a maximum second-harmonic power of 3 mW. The duration of the second harmonic pulse was measured to be 11 ns (full-width at half-maximum) with a photodiode. The laser focus on the sample has been measured to have a diameter ($1/e^2$) of 180 µm based on characterization of a virtual focus and critical pulse energies necessary to melt aluminum and gold films.

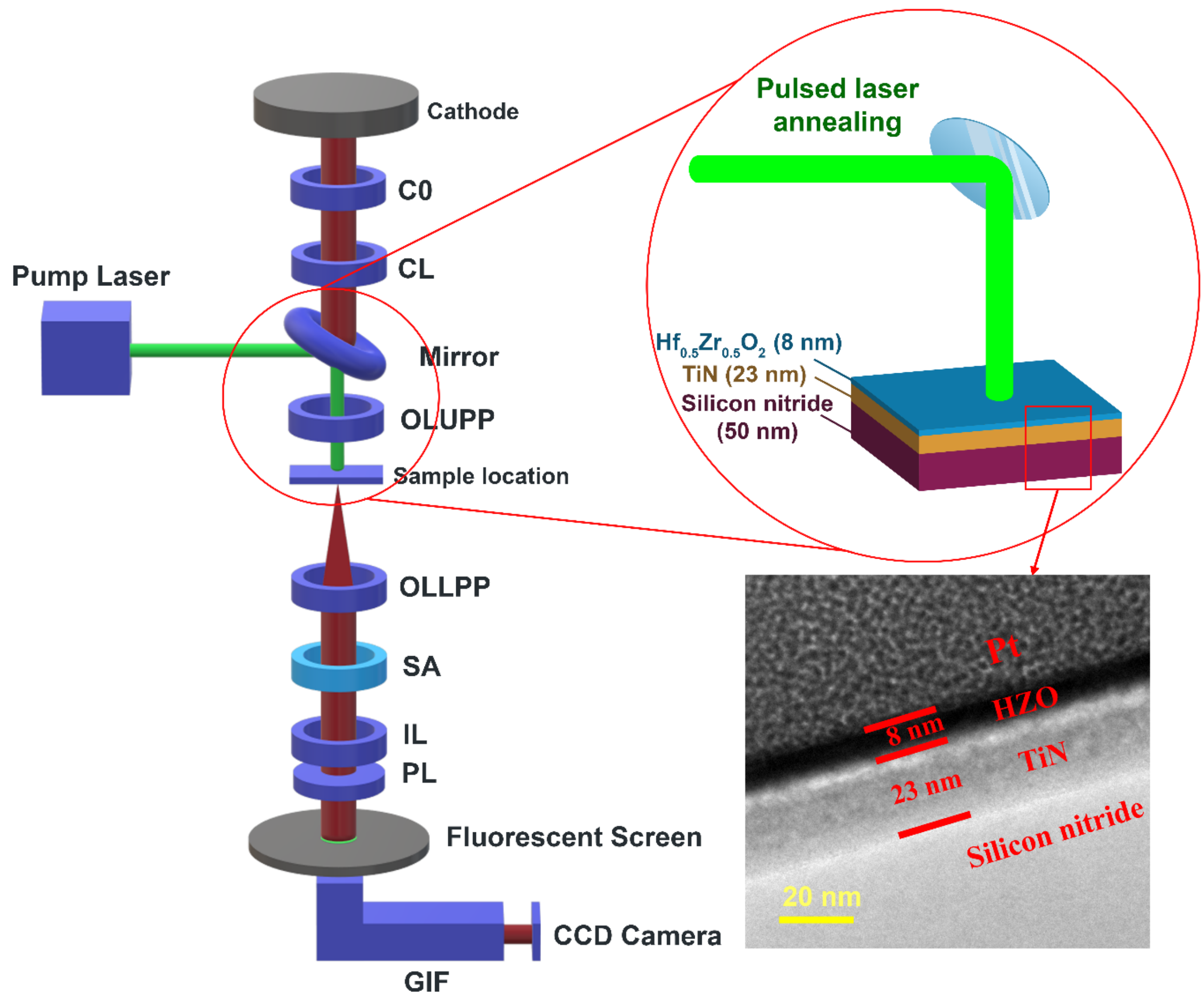

**Figure S2.** The schematic of the modified TEM shown on the left indicates the configuration of the microscope used to carry out in situ PLA experiments. A metallic mirror installed between the condenser lens system (C0 and CL) and the objective lens upper pole piece (OLUPP) allows for illumination of the sample with a laser beam. The electron beam (shown in dark red) passes through a hole in this mirror and scatters from the sample before passing through the objective lens lower pole piece (OLLPP), selected-area aperture (SA), image forming lenses (IL and PL) and GIF to be detected on the post-GIF camera. The inset on the top-right shows a schematic of $Si_3N_4$/TiN/HZO heterostructure relative to the incident laser beam. The lower-right panel shows a TEM image of a lamella cross-sections of the heterostructure prepared by FIB which was used to measure the deposited film thicknesses.

Alt text: The diagram of the microscope configuration shows a laser beam reflecting from a mirror in the column and is directed colinearly onto the sample. An inset of the sample geometry shows the laser beam incident on the top surface of a heterostructure. The heterostructure is made of a 50-nanometer thick silicon nitride membrane that has a 23-nanometer thick titanium nitride film covered by an 8-nanometer thick hafnia-zirconia film.

Two different types of in situ PLA experiments were performed on the samples. First a laser intensity titration experiment was performed, where the same area of the sample was exposed to a series of laser pulses, each subsequent pulse having increased fluence until crystallization was observed. For example, the 8-nm HZO sample was exposed to the laser pulses with 33, 67, and 120 $mJ/cm^2$ before crystallizing at the fluence of 177 $mJ/cm^2$. Then, single-shot PLA experiments were performed where a fresh window in the sample was illuminated by a single laser pulse. Multiple fluences near that found in the intensity titration measurement were attempted to determine the single-shot threshold fluence to induce crystallization. Five titration experiments and more than ten single-shot experiments were performed for each heterostructure design to determine accuracy and reproducibility of determined threshold values.

The annealing process and the crystallization in the HZO samples was investigated through measurements of SAED images of the laser affected region on the sample before and after laser induced crystallization occurs. To be more specific, for the laser titration experiments, we captured the SAED pattern of the targeted region on the sample before being exposed to the laser pulses, and after each pulse. A series of SAED patterns is shown in Figure S3. We used a selected area aperture corresponding to an investigated area on the sample with a diameter of 180 μm. All SAED patterns were captured using 5 cm camera length and an exposure time of 10 seconds.

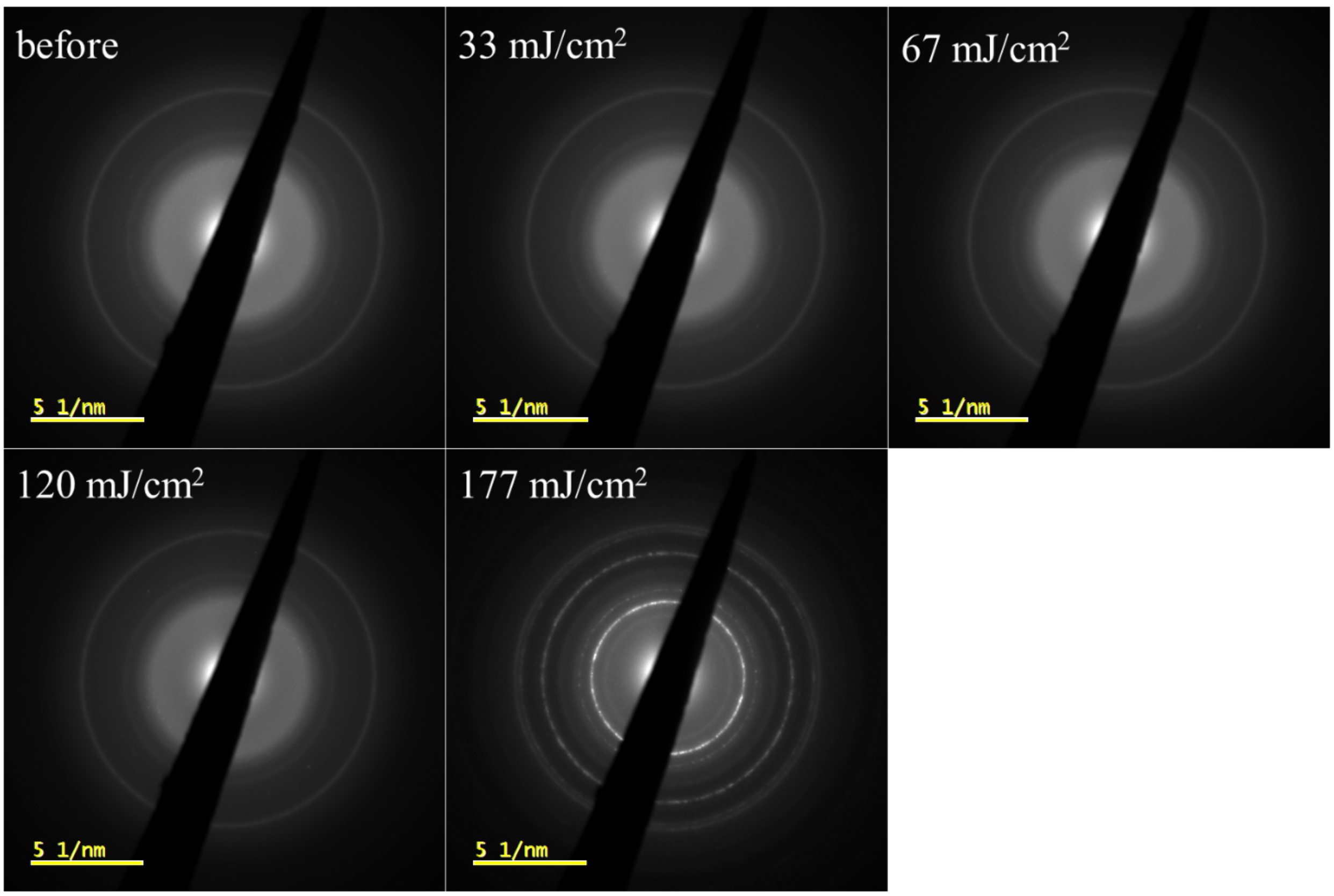

**Figure S3**. SAED images of 8 nm HZO/TiN structure before and after being exposed to the laser pulses with 33 $mJ/cm^2$, 67 $mJ/cm^2$, 120 $mJ/cm^2$, and 177 $mJ/cm^2$ during the titration pulsed laser annealing.

Alt text: The SAED images for the fluences of 33, 67 and 120 millijoules per centimeter squared show low intensity diffuse pattern with a slight diffraction ring at around 7 inverse nanometers. The image of the 177 millijoules per centimeter fluence shows bright and sharp rings showing indicating crystallization of the film.

**2. Selected Area Electron Diffraction (SAED) Quantitative Phase Analysis**

In this study, we aimed to quantitatively assess the influence of HZO film thickness and PLA parameters on the formation of the orthorhombic ferroelectric phase. To achieve this, we developed a detailed analysis of the diffraction intensity radial distribution profiles derived from SAED patterns. The radial profiles were obtained using the *CrysTBox* software[1]. This process involved calculating the radial profiles as the averaged values of the diffraction ring intensities, masking out the beam stop to exclude it from the averaging process, and carrying out a three-step procedure to locate and refine the center of the diffraction pattern. The intensity profile measured from a bare silicon nitride substrate was then subtracted from the data.

**Table S1.** Crystal lattice parameters of the phases considered in SAED QPA

| | **m-HZO** | **oIV-HZO** | **t-HZO** | **TiN** |
|---|---|---|---|---|
| **Compound** | $Hf_{0.5}Zr_{0.5}O_2$ | $Hf_{0.5}Zr_{0.5}O_2$ | $Hf_{0.5}Zr_{0.5}O_2$ | TiN |
| **Space group** | $P2_1/c$ | $Pca2_1$ | $P4_2/nmc$ | $Fm\bar{3}m$ |
| **Symmetry** | monoclinic | orthorhombic | tetragonal | cubic |
| **a, b, c [nm]** | 0.511, 0.518, 0.528 | 0.524, 0.501, 0.505 | 0.356, 0.356, 0.513 | 0.420 |
| **α, β, γ [°]** | 90.0, 99.6, 90.0 | 90, 90, 90 | 90, 90, 90 | 90, 90, 90 |

Alt text: The unit cell parameters of monoclinic, orthorhombic and tetragonal hafnia-zirconia as well as titanium nitride are given.

The positions and intensities of Bragg peaks of each crystalline phase considered in the modelling were then calculated. Table S1 summarizes the crystallographic information used for each phase assumed in computing this information, which was taken from the literature[2, 3]. The associated CIF files that detail atomic coordinates are provided as supplementary information. We then computed the electron diffraction peak intensities using Equations S1 and S2.

$$F_{hkl} = \sum_n f_n(k_{hkl}) e^{2\pi i(hx_n + ky_n + lz_n)} \tag{S1}$$

$$I_{hkl} = LM_{hkl}F_{hkl}^2 \tag{S2}$$

Equation S1 defines the structure factor, $F_{hkl}$, corresponding to Bragg peaks indexed by Miller indices, $hkl$, in terms of the atomic positions in the unit cell, represented by $x_n$, $y_n$, and $z_n$. The term $f_n$ is the electron scattering factor of the $n$-th atom in the unit cell, which was taken from the International Tables for X-ray Crystallography [4]. The intensity for the reflection, $I_{hkl}$, is calculated using Equation S2, where $M_{hkl}$ is the multiplicity factor, and $L$ is the Lorentz factor that we have taken to be proportional to $1/\sin^2\theta_{hkl}$ [4].

Subsequently, a Python-based fitting algorithm was developed to model the SAED radial profiles. A model diffraction pattern composed of multiple phases was represented as weighted sum following:

$$I(k) = BG(k) + \sum_{\alpha} S_{\alpha} \sum_{hkl_{\alpha}} I_{hkl_{\alpha}} f(k, k_{hkl_{\alpha}}, \sigma_{\alpha}) \,, \tag{S3}$$

where $BG(k)$ is a polynomial background function, and the diffraction pattern for each phase, $\alpha$, is the sum of the set of allowed peaks for that phase, $hkl_{\alpha}$, weighted by a scale factor, $S_{\alpha}$. The peak shape function, $f(k, k_{hkl_{\alpha}}, \sigma_{\alpha})$, was assumed to be a gaussian centered at $k_{hkl} = 1/d_{hkl}$ and have a standard deviation of $\sigma_{\alpha}$. The peaks were then scaled by the intensities, $I_{hkl_{\alpha}}$, calculated from Equation S2. The model was constructed using the *lmfit* python library[5] and refinement against experimental data was performed using the associated non-linear least squares minimizer algorithm. In the refinement of the model, only the parameters of the background, and phase-specific parameters $S_{\alpha}$ and $\sigma_{\alpha}$ were varied. As will be demonstrated, this approach allowed us to accurately quantify the phase fractions and explore the effects of varying film thickness and PLA conditions on phase stability. The phase fraction of each HZO phase present in the model, $p_{\alpha_{HZO}}$, was calculated using the equation

$$p_{\alpha_{HZO}} = \frac{S_{\alpha_{HZO}}}{\sum S_{\alpha_{HZO}}}, \tag{S4}$$

where the scale factor of a given HZO phase, $S_{\alpha_{HZO}}$, is normalized by the sum of scale factors determined for all HZO phases present in the model.

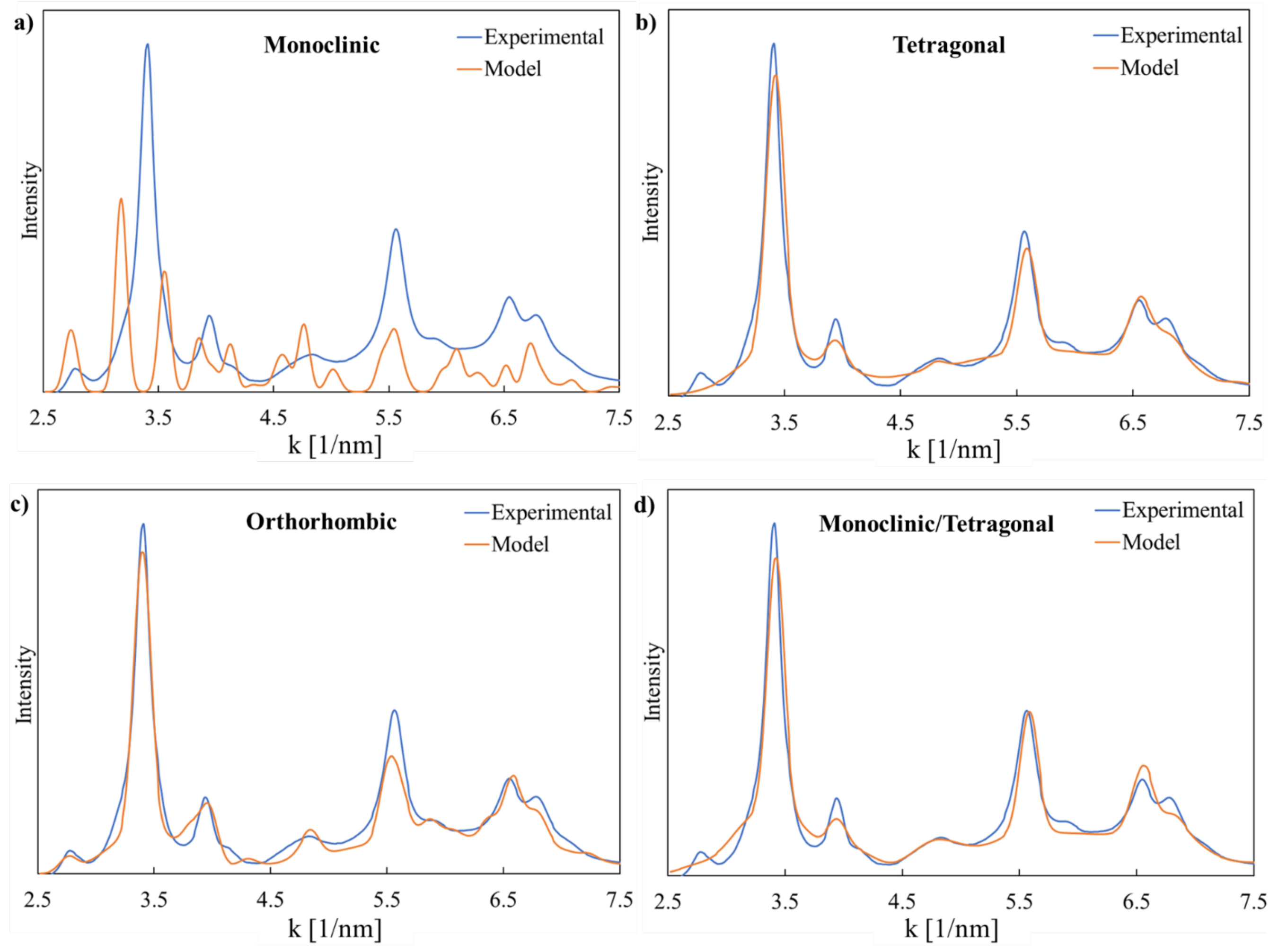

**Figure S4.** The model and experimental electron diffraction intensity radial distribution profiles corresponding to SAED patterns of 8-nm HZO sample after laser titration PLA (a) TiN+m-HZO model ($\chi^2$ = 1.469), (b) TiN+t-HZO model ($\chi^2$ = 0.052), (c), TiN+o-HZO model ($\chi^2$ = 0.051), and (d) TiN+m-HZO+t-HZO model ($\chi^2$ = 0.051)

Alt text: Example fits for different assumed models show that the best fit is obtained for the model that considers a mixture of titanium nitride and orthorhombic hafnia-zircona.

## 3. Supplementary Information on COMSOL Simulations

### 3.1 Simulation methodology

A three-dimensional geometry of the $Si_3N_4$ / TiN / HZO heterostructure was defined consisting of stacked square-planar layers having thicknesses consistent with measured values reported in the manuscript. The effect of laser irradiation was represented by time-dependent heat sources in the TiN and HZO layers having a Gaussian spatial and a flat-top temporal profile. Expressions for the heat sources of each layer, which consider the reflectivity and absorption of each material are given below. Heat was primarily generated in the TiN layer as it has an absorption coefficient of 4.6 x $10^5$ $cm^{-1}$ at 532 nm [6], while that of hafnia is only 0.64 $cm^{-1}$ [7] and that of $Si_3N_4$ is effectively zero.

A 2D fine mesh with element sizes between 5.4 um and 43.2 um in the plane was swept in the z-axis (surface normal direction) and found to be suitable to capture the temperature evolution in the system. The simulation geometry relative to the size of the heat affected zone is shown in Figure S5.

Time-dependent simulations of the temperature evolution in the heterostructures were then carried out using the COMSOL heat transfer in solids module. During the simulation, the temperature around the edges of the layers was fixed at 300 K, and the top and bottom surfaces were isolated to simulate the vacuum environment inside the TEM. Important physical constants, such as the heat capacity and thermal conductivity have not been measured for HZO. Therefore, we conducted two sets of simulations assuming the HZO layer had the properties of either pure hafnia or pure zirconia to define a reasonable range for the temperature evolution. The physical constants for the heat capacity, thermal conductivity, and density of each material used in our simulations are given in Supplementary Table S2. In addition, we accounted for the partial melting of the silicon nitride layer by including a solid-to-liquid reversible phase transformation of this material at 1900 °C, which was modelled using the apparent heat capacity method implemented in COMSOL. The values for the temperature transition interval and latent heat for this transformation are also given in Table S2.

### 3.2 Definition of Heat Source

The power of the heat sources as function of position within the HZO and TiN layers was defined by

$$P_{HZO}(x,y,z) = \frac{(1-R_{HZO})Ep*\alpha_{HZO}*e^{-\alpha_{HZO}*Z}*e^{\frac{-((x-(1.5*D))^2+(y-(1.5*D))^2)}{2\sigma^2}}}{2\pi p_w \sigma^2} \quad \text{(S5)}$$

$$P_{TiN}(x,y,z) = \frac{(1-R_{TiN})*(1-R_{HZO})Ep*\alpha_{TiN}*e^{-\alpha_{HZO}*t_{HZO}}*e^{(-\alpha_{TiN}*(Z-t_{HZO}))}*e^{\frac{-((x-(1.5*D))^2+(y-(1.5*D))^2)}{2\sigma^2}}}{2\pi p_w \sigma^2} \quad \text{(S6)}$$

The variables represent the following parameters:

| Variable | Parameter [units] |
|---|---|
| $R_{TiN}$ | Reflectance of TiN |
| $R_{HZO}$ | Reflectance of HZO |
| $\alpha_{HZO}$ | HZO Absorptivity [$m^{-1}$] |
| $\alpha_{TiN}$ | TiN Absorptivity [$m^{-1}$] |
| $p_w$ | Pulse Duration [s] |
| $\sigma$ | Gaussian Laser spot size [m] |
| $E_p$ | Pulse Energy [J] |
| $D$ | Laser spot diameter = 4$\sigma$ [m] |
| $t_{HZO}$ | Thickness of HZO layer [m] |

The temporal dependance of these heat sources was a square pulse with a 11 ns duration equal to the incident laser pulse. Physical parameters of the materials assumed in the simulations is detailed in the following table.

**Table S2.** Thermodynamic parameters of materials assumed in COMSOL simulations

| | Density<br>[ kg / m$^3$ ] | Heat Capacity<br>[ J / kg K ] | Thermal Conductivity<br>[ W / m K ] | Reflectivity<br>- | Absorptivity<br>[1/m] | Latent Heat<br>[ kJ / kg ] | Phase change temperature<br>[ K ] | Transition interval<br>[ K ] |
|---|---|---|---|---|---|---|---|---|
| **$Si_3N_4$** | 3190 [8] | 1024 @ 773 K<br>1252 @ 1400K [9] | 2.69 [COMSOL]<br>30 [10] | - | - | 5914 [NIST] | 2173 [NIST] | 10 [NIST] |
| **$HfO_2$** | 7265 [11] | 332 @ 1000 K [12] | 2.39 @ 1500 K [13] | 0.1321 | 64 | - | - | - |
| **$ZrO_2$** | 5870 [14] | 568 @ 1000K [12],<br>644 @ 1400K [9] | 2.11 @1400 K [15] | 0.1321 | 64 | - | - | - |
| **TiN** | 5440 | 658 [COMSOL] | 3.79-3.89 | 0.4392 | 4.6126E7 | - | - | - |

Alt text: Physical and thermal parameters for silicon nitride, hafnia, zirconia, and titanium nitride are given.

Simulations were also carried out without considering partial melting of silicon nitride. In this case the peak temperature was found to reach more than 4000 °C, which is far beyond the melting point of this material and does not match with our observations.

Further simulations were performed to evaluate the sensitivity of the thermal response to HZO thickness and assumed HZO thermodynamic properties. As shown in Figure S6, As shown in Figure S6, the simulated temperature evolution is very similar for 7-nm, 8-nm, and 15-nm HZO layers, as well as when hafnia-like or zirconia-like thermodynamic properties are assumed. While the peak temperature achieved within the first few nanoseconds was found to differ by 50 °C, in all cases partial incomplete melting of the silicon nitride membrane began at 6 ns and invariably limited the temperature of the HZO film near 1900 °C. Therefore, these simulations suggest that the $Si_3N_4$ film in fact plays the dominant role of an effective heat sink at 1900 ºC in determining the temperature dynamics of the sample, whereas HZO thickness has only a minor effect on the simulated peak temperature.

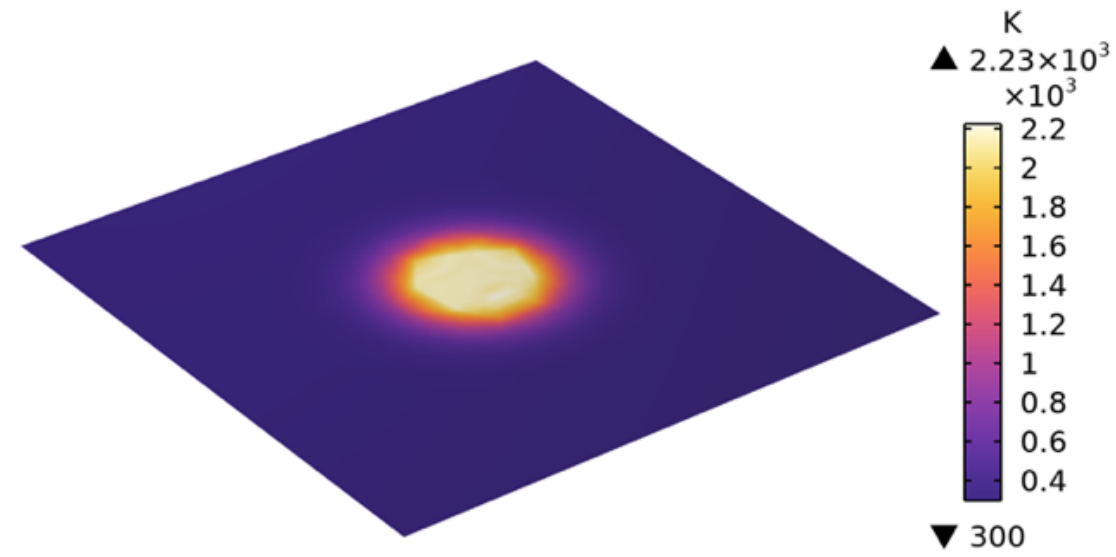

**Figure S5.** COMSOL simulation geometry relative to the size of the heat affected zone of the sample heterostructure.

Alt text: Geometry of the COMSOL simulation shows that the heat affected area is 3 times smaller than the sample surface.

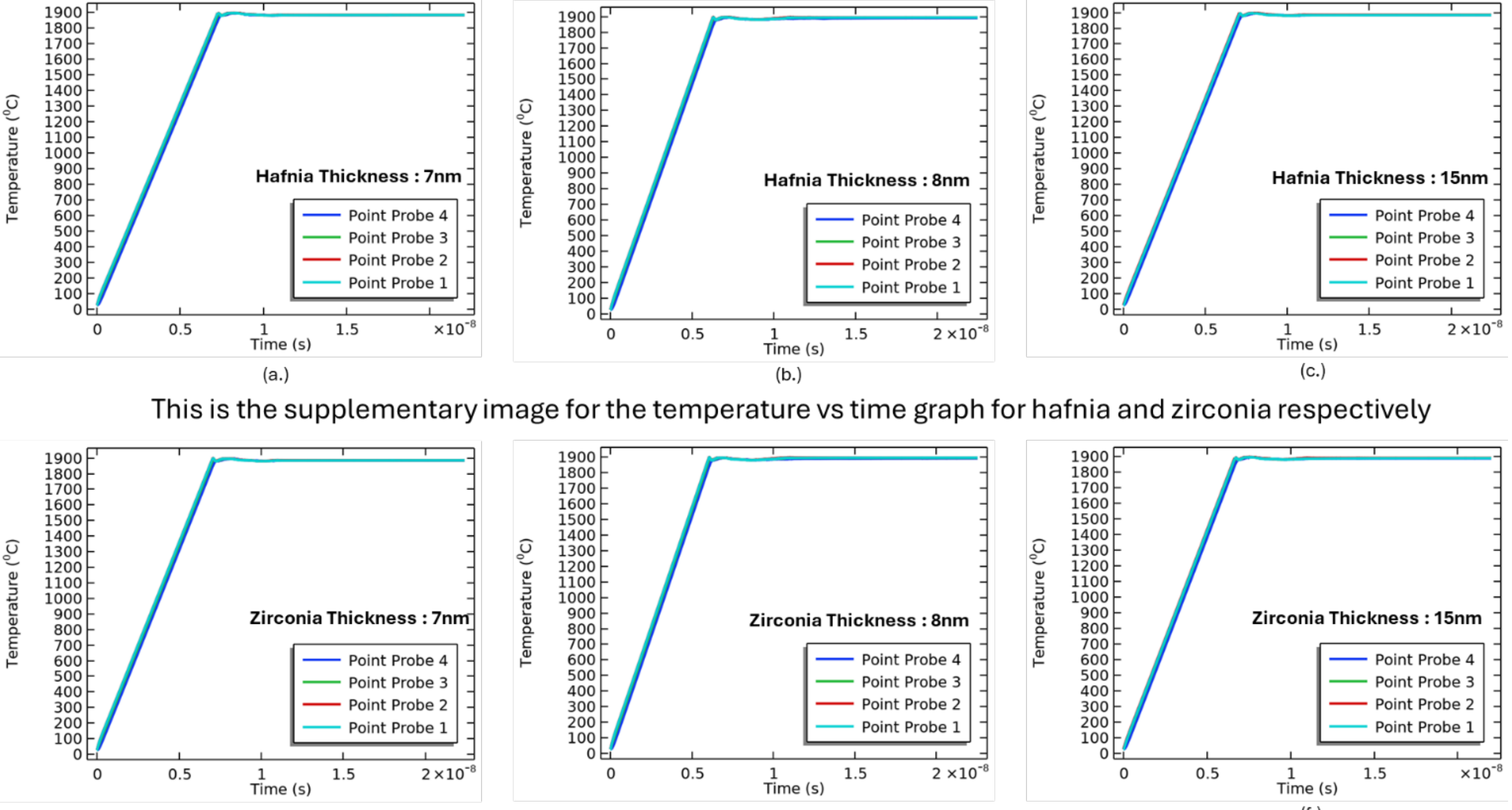

**Figure S6.** Sensitivity analysis of the simulated temperature evolution for multilayer heterostructures with 7-, 8-, and 15-nm HZO layers and with HZO thermodynamic properties approximated by hafnia or zirconia. The positions of temperature point probes 1-4 are the same as indicated in Figure 4a.

Alt Text: The same temperature evolution is found for all simulations regardless of film thickness and assuming different approximations of the film as having the properties of either hafnia or zirconia.

## 4. References

1. M. Klinger, *Journal of Applied Crystallography*, **50**, 1226-1234 (2017).
2. J. Muller, T. S. Boscke, U. Schroder, S. Mueller, D. Brauhaus, U. Bottger, L. Frey and T. Mikolajick, *Nano letters*, **12**, 4318-4323 (2012).
3. R. Materlik, C. Künneth and A. Kersch, *Journal of Applied Physics*, **117**, 134109-134101-134109-134115 (2015).
4. E. Prince and P. Boggs, **C, Chapter 8.1**, 678–688 (2006).
5. M. Newville, T. Stensitzki, D. B. Allen, M. Rawlik, A. Ingargiola and A. Nelson, *Astrophysics Source Code Library*, ascl: 1606.1014 (2016).
6. J. Pflüger, J. Fink, W. Weber, K. Bohnen and G. Crecelius, *Physical Review B*, **30**, 1155-1163 (1984).
7. T. J. Bright, J. I. Watjen, Z. Zhang, C. Muratore and A. A. Voevodin, *Thin Solid Films*, **520**, 6793-6802 (2012).
8. H. S. Dow, W. S. Kim and J. W. Lee, *AIP Advances*, **7** (2017).

9. M. W. Chase, *Journal of physical and chemical reference data*, **28**, 1951 (1998).
10. AZM website, Thermal properties of Silicon nitride https://www.azom.com/article.aspx?ArticleID=5067).
11. J. Gong, X. Liu, L. Yang, A. Sulyok, Z. Baji, V. Kis, K. Tőkési, R. Zeng, G. Fang and J. Gong, *Journal of Alloys and Compounds*, **1005**, 175744 (2024).
12. X. Luo, W. Zhou, S. V. Ushakov, A. Navrotsky and A. A. Demkov, *Physical Review B—Condensed Matter and Materials Physics*, **80**, 134119 (2009).
13. X. Xiang, H. Fan and Y. Zhou, *Journal of Applied Physics*, **135** (2024).
14. A. Opalinska, I. Malka, W. Dzwolak, T. Chudoba, A. Presz and W. Lojkowski, *Beilstein Journal of Nanotechnology*, **6**, 27-35 (2015).
15. L. Momenzadeh, I. V. Belova and G. E. Murch, *Computational Materials Science*, **176**, 109522 (2020).